# Relativity and Aether Theory a Crucial Distinction

Joseph Levy

4 Square Anatole France, 91250, St Germain-lès-Corbeil, France
E-mail: levy.joseph@orange.fr

We study the case of two rockets which meet at a point O of an 'inertial co-ordinate system' $S$, and are scheduled to move at constant speed, in opposite directions, toward two targets placed at equal distances from point O. At the instant they meet, the clocks inside the rockets are set to zero. When they reach the targets the rockets meet two clocks A and B whose *reading* is identical. This question which was tackled in ref [1] is studied here in depth. Assuming the existence of a preferred aether frame $S_0$ in which the one-way speed of light is isotropic, and the anisotropy of this speed in the other frames, we show that, if the equal reading of the clocks A and B results from an exact synchronization, the clocks inside the rockets will display different readings when they reach A and B in contradiction with the relativity principle. Conversely, if the clocks A and B, which display an equal reading, have been synchronized by means of the Einstein-Poincaré procedure, the inboard clocks will also display the same reading, a fact which seems in agreement with the relativity principle. But this synchronization method presupposes the invariance of the one-way speed of light, in contradiction with the assumptions made, and, therefore, introduces a measurement error. This demonstrates that if we assume the existence of an aether frame, the *apparent* relativity principle is not a fundamental principle; it depends on an arbitrary synchronization. In any case, this is an example of an experimental measurement which can be explained by aether theory without the assumption of the invariance of the one-way speed of light in all 'inertial frames'.

---

Version supplemented by additional information and other references

## 1. INTRODUCTION

A number of arguments today lend support to the existence of a preferred aether frame in which the one-way speed of light is isotropic [1] and to the anisotropy of this speed in the other frames, and it is of the utmost importance to know whether such a preferred frame is compatible with the application of the relativity principle in the physical world. Physicists remain divided about this question. Einstein was convinced that the existence of a preferred frame is at variance with relativity. In the original formulation of his theory [2], he definitely regarded the existence of aether as superfluous. Later he changed his mind in order to formulate the theory of general relativity. But, the aether of Einstein is not associated with a preferred frame. In his little book "Sidelights on relativity" [3], he expressed his views in the following terms:

> "..according to the theory of general relativity, space is endowed with physical qualities. In this sense, therefore there exists an aether… But this aether may not be thought of as endowed with the quality of ponderable media, as consisting of parts which may be tracked through time. The idea of motion may not be applied to it".

On the contrary, Poincaré acknowledged the Lorentz assumptions which assume the existence of a preferred aether frame and in which length contraction and clock retardation are real processes depending on the velocity of the rods and clocks relative to the aether frame. The agreement of Poincaré with the approach of Lorentz is expressed in the following sentence:

> "The results I have obtained agree with those of Mr. Lorentz in all important points. I was led to complete and modify them in a few points of detail" [4].

His belief in the aether was expressed in the citations that follow:

> Does aether really exist? The reason why we believe in aether is simple. If light comes from a distant star and takes many years to reach us, it is (during its travel) no longer on the star, but not yet near the Earth. Nevertheless, it must be somewhere and supported by a material medium; (La science et l'hypothèse chapter 10 p 180 of the French edition "Les theories de la physique moderne" [5]).

And:

> "Let us remark that an isolated electron moving through the aether, generates an electric current, that is to say an electromagnetic field. This field corresponds to a certain quantity of energy localized in the aether rather than in the electron" [6].

But, at the same time, Poincaré acknowledged the relativity principle, as the following sentence shows:

"It seems that the impossibility of observing the absolute motion of the Earth is a general law of nature. We are naturally inclined to admit this law that we shall call the relativity postulate and to admit it without restriction" [7].

In this text, we propose to check these different opinions starting from a simple experimental test.

**11. OVERVIEW OF THE PROBLEM**
Let us consider two rockets moving uniformly in opposite directions along a straight line of an 'inertial co-ordinate system'[1] *S*. At the initial instant (0) the rockets meet at a point O and their clocks are set to zero. The rockets are scheduled[2] to move at constant speed toward two points A and B placed at equal distances from point O where they meet two clocks whose reading is identical (see Fig 1). When the rockets reach points A and B, their inboard clocks are stopped and then compared. There is neither acceleration nor deceleration during the process.

According to Einstein's special relativity, the inboard clocks should display the same reading when they stop; indeed, since the speed of light is regarded as isotropic in all 'inertial' frame, it is assumed that no obstacles are opposed so that an exact synchronization is carried out. Therefore the equal reading displayed by the clocks A and B is regarded as the real time. Due to the complete symmetry of the transit of the two rockets, their inboard clocks must display the same reading, which is equal to the reading of the clocks A and B multiplied by the $1/\gamma$ factor. This is a condition so that the relativity principle is obeyed.

As we shall see, a completely different explanation is provided by aether theory.
According to Poincaré's theory, as we have seen, there is no assumed incompatibility between the existence of a privilege frame and the principle of relativity. Is this really the case? This test will enable us to answer this question in the following chapters. (We must bear in mind that, in aether theory, clock retardation is defined with respect to the privileged aether frame which is represented here by the co-ordinate system $S_0$).

---

[1] Let us remember that perfect inertial frames don't exist in the physical world. The concept must be regarded as a limit case, which real frames approximate more or less.
[2] Note that, as we shall see in the following chapters, even though the rockets are scheduled to move symmetrically, the symmetry will be only apparent if the clocks used to measure the speeds are affected by a synchronism discrepancy effect.

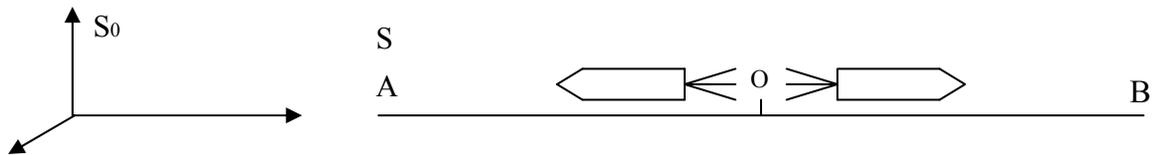

Fig 1: The two rockets are scheduled to move, at constant speed toward two clocks placed in A and B at equal distances from point O. When they reach points A and B, the reading of these clocks is identical.

**111. MEASUREMENT AND CLOCK SYNCHRONIZATION**
Assuming the existence of a preferred aether frame, this issue needs to be considered successively from two different points of view.

111. 1 *<u>The first point of view presumes that one can exactly measure the transit time of the rockets from point O to points A and B .</u>*
This implies that the identical reading $\tau$ of the clocks placed at points A and B, when the rockets reach them, which was assumed by definition, translates the identity of the real time. This fact implies a perfect synchronization of the clocks placed at points A and B with the clock placed at point O.
(Yet we know that synchronizing clocks exactly is not an easy process [8]).

In any case an exact synchronization can be considered, even if we cannot do it exactly nowadays and it is justified to estimate the implications of such a procedure[3]
Let us therefore first suppose, for our purpose, that this exact synchronization of clocks has been carried out.
Assuming in agreement with aether theory that clock retardation results from the motion of the rockets with respect to the aether frame, the resolution of the problem is easy. Insofar as the rockets do not have the same speed with respect to the aether frame, the slowing down of their inboard clocks will be different and they will display different readings. (Only if the co-ordinate system $S$ was at rest with respect to the aether frame, the clocks inside the rockets would display the same reading

---

[3] Notice that an accurate synchronization of clocks is not impossible knowing that different experiments and astronomical observations have permitted estimation of the absolute speed of the Earth frame and, therefore, of the magnitude of the one-way speed of light [1]. Most probably, in the near future, a more accurate determination of this speed will enable us to synchronize the clocks almost exactly

irrespective of their direction of motion). Of course, this would inform us whether $S$ is at rest or in motion relative to $S_0$, in contradiction with the principle of relativity.

Therefore insofar as the rockets' speeds are determined exactly, Poincaré's relativity principle is shown to be at variance with the existence of a preferred aether frame.

111. 2 *<u>We shall now study what happens when the transit time of the rockets is measured, in the co-ordinate system S, with clocks synchronized by means of the usual Einstein-Poincaré procedure.</u>*

In order to synchronize the clocks placed at points A and B, we shall make use of the Einstein-Poincaré synchronization procedure (E. P synchronization) which assumes that the speed of light is equal to C in all inertial frames. To this end, we send a light signal at time $t_0 = 0$ from clock O to clock A (or B). After reflection, the signal returns to O. The clock is supposed to be synchronous with clock O if, at the instant of reflection, it displays the reading $t = T/2$, where T is the reading displayed by clock O at the instant when the signal comes back to it.

Insofar as the one-way speed of light is not isotropic in co-ordinate systems which are not at rest with respect to the aether frame, the use of this method introduces an unavoidable systematic error that must be corrected, as we shall see below.

For convenience, we shall assume that the segment AB is aligned and moves along the $x$-axis of the co-ordinate system $S_0$ which is at rest with respect to the aether frame. Let us refer to the length of the segment AB when it is at rest in the aether frame as $2\ell$. Since it is moving with respect to the aether frame at speed $v$, half of its length (measured with a non-contracted standard) will be $\ell\sqrt{1-v^2/C^2}$ where $C$ is the speed of light in the aether frame.

Actually, according to the aether theory considered in this text, the real speed of light relative to the co-ordinate system $S$ along the direction A→B is equal to $C - v$, and in the opposite direction to $C + v$. Even though the magnitude of $v$ is not exactly known, this assumption will be helpful for our purpose. (These formulas were the expressions used by Lorentz to explain the Michelson experiment). As we shall see in appendix 2, for the present case, and in ref [1], for the general cases, only speeds whose measurements are altered by the systematic measurement distortions obey the relativistic law of composition of velocities.

The real time needed by the light signal to travel from point O to point B is therefore:

$$t_{rB} = \frac{\ell\sqrt{1-v^2/C^2}}{C-v}$$

(where the suffix r means real)

$t_{rB}$ is the time that, in the absence of clock retardation, the clock placed in B and exactly synchronized with clock O would display when the signal which starts at instant zero from point O reaches point B.

Taking account of clock retardation in *S*, the clock reading in the absence of synchronism discrepancy effect would be:

$$t_{rB}\sqrt{1-v^2/C^2} = \ell\frac{(1-v^2/C^2)}{C-v}$$

But, what we measure by means of the E. P synchronization procedure is half the reading displayed by clock O at the instant when the signal returns to it.

With clocks not slowed down by motion, the *apparent* time needed by the light signal to travel from point O to point B would be:

$$t_{Bapp} = 1/2\ell\sqrt{1-v^2/C^2}(\frac{1}{C-v}+\frac{1}{C+v}) = \frac{\ell}{C\sqrt{1-v^2/C^2}}$$

And the reading displayed by clock B when one takes account of clock retardation is:

$$t_{Bapp}\sqrt{1-v^2/C^2} = \ell/C$$

(this expression is equal to the reading $t = T/2$ defined above)

(We can see that, contrary to special relativity, aether theory does not consider the ratio $\ell/C$ as the real time of light transit from O to B).

Thus, taking account of clock retardation in *S*, the synchronism discrepancy of clock B with respect to clock O is:

$$\Delta = (\frac{\ell\sqrt{1-v^2/C^2}}{C-v} - \frac{\ell}{C\sqrt{1-v^2/C^2}})\sqrt{1-v^2/C^2}) = \frac{v\ell}{C^2}$$

(We can see that the apparent time is shorter than the real time.)

We shall now determine the synchronism discrepancy of clock A with respect to clock O. We can easily anticipate that it will be equal to $-\Delta$, but, even so, the calculation deserves to be done.

The real time needed by the light signal to travel from point O to point A is:

$$t_{rA} = \frac{\ell\sqrt{1-v^2/C^2}}{C+v}$$

It is the time that, in the absence of clock retardation, the clock placed in A and exactly synchronized with clock O would display when the signal reaches point A.

Taking account of clock retardation in the co-ordinate system *S* and using an exact synchronization procedure, the reading displayed by clock A would be:

$$t_{rA}\sqrt{1-v^2/C^2} = \ell\frac{(1-v^2/C^2)}{C+v}$$

But, what we measure by means of the E. P synchronization procedure is half the reading displayed by clock O at the instant when the signal returns to it.
With clocks not slowed down by motion, the *apparent* time needed by the signal to travel from point O to point A would be:

$$t_{Aapp} = 1/2\ell\sqrt{1-v^2/C^2}\left(\frac{1}{C+v} + \frac{1}{C-v}\right) = \frac{\ell}{C\sqrt{1-v^2/C^2}}$$

Therefore, the reading displayed by clock A when one takes account of clock retardation is:

$$t_{Aapp}\sqrt{1-v^2/C^2} = \ell/C$$

(it is the same as the reading displayed by clock B).
Thus, taking account of clock retardation, the synchronism discrepancy of clock A with respect to clock O is:

$$\Delta' = -\Delta = \left(\frac{\ell\sqrt{1-v^2/C^2}}{C+v} - \frac{\ell}{C\sqrt{1-v^2/C^2}}\right)\sqrt{1-v^2/C^2} = -\frac{v\ell}{C^2}$$

We note that, contrary to clock B the apparent time given by clock A is longer than the real time.

Let us now study the effect of the synchronism discrepancy on the clocks placed inside the rockets.

In the experiment, the *apparent* transit times of the rockets relative to point O, measured by an observer at rest relative to *S* using the E. P synchronization procedure, are assumed to be identical by definition; therefore, when the rockets reach points A and B, the clocks A and B will display the same reading $\tau$.
But due to the synchronism discrepancy effect this reading is erroneous and must be corrected and, as we shall see in the text that follows the real transit times of the rockets are in fact different, and, of course, their real speed also differ.
In fact, in the absence of synchronism error, the reading of clock B would have been:

$$\tau + \Delta = \tau + \frac{v\ell}{C^2}$$

And the reading of clock A:

$$\tau - \Delta = \tau - \frac{v\ell}{C^2}$$

Let us now determine the real transit times $t_{0A}$ and $t_{0B}$ that would be displayed by clocks attached to frame $S_0$ when the rockets reach points A and B. We have:

$$\tau + \Delta = t_{0B}\sqrt{1-v^2/C^2}$$

and

$$\tau - \Delta = t_{0A}\sqrt{1 - v^2/C^2}$$

Thus:

$$t_{0B} = \frac{\tau + v\ell/C^2}{\sqrt{1 - v^2/C^2}} \tag{1}$$

And

$$t_{0A} = \frac{\tau - v\ell/C^2}{\sqrt{1 - v^2/C^2}} \tag{2}$$

We note that these expressions assume the same mathematical form as the conventional transformations, yet, as we saw, since they have been measured with clocks E. P synchronized and contracted meter sticks, $\tau$ and $\ell$ are not the real space and time co-ordinates of the points A and B when the rockets reach these points (see Ref [1]).

Readings displayed by the clocks inside the rockets.

Since, according to our initial conditions, the *apparent* transit times of the rockets in *S*, measured with clocks E. P synchronized, are identical, and equal to $\tau$, the apparent speeds will be $v_{app} = \ell/\tau$ in both sides, (where $2\ell$ is the *apparent* length of AB in *S*, measured with a contracted standard). Yet the apparent time corresponds to two different real times $t_{0A}$ and $t_{0B}$ and therefore to two different real speeds *v'* and *v"*.

Using these values we can determine the *apparent* transit times displayed by the clocks in the rockets' frames $T_A$ and $T_B$ and therefore we shall see that they are identical, although the real times given by formulas (1) and (2) are not.

Taking account of clock retardation, the clock present inside the rocket travelling toward point B, at the instant when it reaches this point, displays the reading:

$$T_B = t_{0B}\sqrt{1 - (v+v')^2/C^2}$$
$$= \frac{\tau + v\ell/C^2}{\sqrt{1 - v^2/C^2}}\sqrt{1 - (v+v')^2/C^2}$$

Where *v'* is the real speed relative to point O of the rocket travelling toward point B.
And the clock of the rocket travelling toward point A will display:

$$T_A = t_{0A}\sqrt{1 - (v-v")^2/C^2}$$
$$= \frac{\tau - v\ell/C^2}{\sqrt{1 - v^2/C^2}}\sqrt{1 - (v-v")^2/C^2}$$

Where *v"* is the real speed relative to point O of the rocket travelling toward A.

We easily verify that

$$v' = \frac{\ell\sqrt{1-v^2/C^2}}{t_{0B}} = \frac{\ell(1-v^2/C^2)}{\tau + v\ell/C^2} \quad (3)$$

and

$$v'' = \frac{\ell\sqrt{1-v^2/C^2}}{t_{0A}} = \frac{\ell(1-v^2/C^2)}{\tau - v\ell/C^2} \quad (4)$$

Replacing $v'$ and $v''$ with their values in $T_A$ and $T_B$ we remark that $T_A$ and $T_B$ are identical. We find:

$$T_A^2 = T_B^2 = \gamma^2(\tau^2 - \ell^2/C^2 - v^2\tau^2/C^2 + v^2\ell^2/C^4)$$

$$= \tau^2 - \ell^2/C^2$$

(See the demonstration in appendix 1)

For values of $\ell/\tau \ll C$ we obtain:

$$T_A = T_B \approx \tau(1 - 1/2\frac{\ell^2}{C^2\tau^2}) = \tau(1 - 1/2\frac{v_{app}^2}{C^2}) \quad (5)$$

For the usual transits whose speed is low compared to the speed of light, this expression approximates $\tau$, a result which highlights the equivalence of the slow clock transport synchronization procedure and the Einstein-Poincaré method, and provides a key to understand the GPS measurements.

### 4. CONCLUSION

This result is very enlightening. It demonstrates that, if we assume the existence of an aether frame and if the measurements of the rockets' transit times from O to A and B, by the observer at rest in *S*, are exactly determined and are found identical, the clocks inside the rockets will display different readings when they reach points A and B. Therefore the relativity principle does not apply with real speeds.

Conversely, if one uses the Einstein-Poincaré procedure in *S* to determine the 'transit times' (and therefore the 'speeds') and if the measurement yields the same clock reading $\tau$ in both sides, then the clocks inside the rockets will also display the same reading when they reach points A and B. This result is due to the systematic inevitable error made when, assuming the isotropy of the one-way speed of light in all 'inertial' frames, one relies on this synchronization procedure with light signals. Therefore the study also verifies the agreement of the slow clock transport synchronization with the E. P procedure in accord with the conclusion of several authors[8][4].

Therefore, assuming the existence of a preferred aether frame implies that the relativity principle is not a fundamental postulate of physics; it depends on arbitrary synchronization procedures.

We emphasize that, although in this experiment, the use of the synchronization procedures mentioned above, to measure the transit times in *S*, make sure that the clocks inside the rockets will display the same reading when they reach points A and B, (in agreement with what special relativity asserts), the interpretation of this fact by aether theory is completely different. In particular, this result has been obtained without assuming the isotropy of the one-way speed of light in the co-ordinate system *S*, a fact which should result in significant consequences for the understanding of physics.

## 5. APPENDIX 1
*Identical readings displayed by the clocks present inside the rockets when the Einstein-Poincaré synchronization procedure is used.*

We have

$$T_B^2 = \gamma^2(\tau + v\ell/C^2)^2[1 - \frac{v^2}{C^2} - 2\frac{vv'}{C^2} - \frac{v'^2}{C^2}]$$

Where $\gamma = (1 - v^2/C^2)^{-1/2}$

Replacing *v'* by its value given in (3), we obtain:

$$T_B^2 = \gamma^2(\tau + \frac{v\ell}{C^2})^2[1 - \frac{v^2}{C^2} - \frac{2v\ell(1-v^2/C^2)}{C^2(\tau + v\ell/C^2)} - \frac{\ell^2(1-v^2/C^2)^2}{C^2(\tau + v\ell/C^2)^2}]$$

$$= \gamma^2[(\tau + \frac{v\ell}{C^2})^2 - \frac{v^2}{C^2}(\tau + \frac{v\ell}{C^2})^2 - \frac{2v\ell}{C^2}(1 - v^2/C^2)(\tau + v\ell/C^2) - \frac{\ell^2}{C^2}(1 - v^2/C^2)^2]$$

$$= \gamma^2(\tau^2 - \ell^2/C^2 - v^2\tau^2/C^2 + v^2\ell^2/C^4)$$

---

[4] A lively debate took place in recent years among physicists about the validity of the relatvity principle. The experiment of Hafele and Keating [9] was presented by the authors as a decisive argument in its favour. Yet the interpretation of the experiment was severely criticized by Kelly [10] and Essen [11]. More recent experiments (including GPS measurements [12, 13]) supported the conclusions of Hafele and Keating. Aether theory provides a key to account for the experimental tests. As shown in this text, the relativity principle seems to apply only with physical data resulting from the measurement distortions. It does not apply any more when the distortions are corrected (see section 111. 2 and the appendixes).

Therefore

$$T_B^2 = \tau^2 - \ell^2/C^2$$

$$T_A^2 = \gamma^2(\tau - v\ell/C^2)^2[1 - \frac{v^2}{C^2} + 2\frac{vv''}{C^2} - \frac{v''^2}{C^2}]$$

Replacing *v''* by its value given in (4), we obtain:

$$T_A^2 = \gamma^2(\tau - \frac{v\ell}{C^2})^2[1 - \frac{v^2}{C^2} + \frac{2v\ell(1-v^2/C^2)}{C^2(\tau - v\ell/C^2)} - \frac{\ell^2(1-v^2/C^2)^2}{C^2(\tau - v\ell/C^2)^2}]$$

$$= \gamma^2[(\tau - \frac{v\ell}{C^2})^2 - \frac{v^2}{C^2}(\tau - \frac{v\ell}{C^2})^2 + \frac{2v\ell}{C^2}(1-v^2/C^2)(\tau - v\ell/C^2) - \frac{\ell^2}{C^2}(1-v^2/C^2)^2]$$

$$= \gamma^2(\tau^2 - \ell^2/C^2 - v^2\tau^2/C^2 + v^2\ell^2/C^4)$$

Therefore

$$T_A^2 = \tau^2 - \ell^2/C^2$$

## 6. APPENDIX 2
### *Composition of velocities law for apparent speeds*
According to the aether theory referred to in this text, speeds are simply additive. The relativistic composition of velocities law results from the measurement distortions caused by length contraction, clock retardation and unreliable clock synchronization, as the following demonstration will show.

We start from the Galilean law $V_B = v + v'$ and $V_A = v - v''$, where $V_B$ and $V_A$ refer to the real speeds of the rockets with respect to the aether frame.

From formulas (1) and (2) we have:

$$\tau = t_{0B}\sqrt{1-v^2/C^2} - v\ell/C^2$$
$$= t_{0A}\sqrt{1-v^2/C^2} + v\ell/C^2$$

$\tau$ is the apparent transit time of the rockets measured in *S* with the clocks A and B (E. P synchronized).

Since OA and OB are measured with a contracted standard they are found equal to $\ell$ although their real length is $\ell\sqrt{1-v^2/C^2}$.

The apparent speed (relative to point O) of the rocket travelling toward B is therefore:

$$v'_{app} = \frac{\ell}{\tau} = \frac{\ell}{t_{OB}\sqrt{1 - v^2/C^2} - v\ell/C^2}$$

We note that by definition $v'_{app} = v''_{app}$ where $v''_{app}$ refers to the apparent speed (relative to point O) of the rocket travelling toward A; a fact which can be easily verified.

(Therefore we will refer to the apparent speed in both directions as $v_{app}$.)

Replacing $t_{0B}$ by its value $\dfrac{\ell\sqrt{1 - v^2/C^2}}{V_B - v}$

we find: $v'_{app} = \dfrac{\ell}{\tau} = \dfrac{\ell}{\dfrac{\ell(1 - v^2/C^2)}{V_B - v} - v\ell/C^2}$

$$= \frac{V_B - v}{1 - \dfrac{v^2}{C^2} - \dfrac{v(V_B - v)}{C^2}} = \frac{V_B - v}{1 - \dfrac{vV_B}{C^2}}$$

This result shows decisively that the relativistic composition of velocities law applies to apparent speeds and not to the real speeds which as we saw are simply additive.

The conclusions drawn in this article are identical to those which were expressed in the previous version submitted to arXiv under the reference *Physics/*0610067. We have only given further explanations and added other references.

**<u>REFERENCES</u>**